\documentclass[epj]{svjour}
\usepackage{amsmath}
\usepackage{amsfonts}
\usepackage{amssymb}
\usepackage{graphics,graphicx}
\usepackage{graphics}
\begin{document}
\title{Braneworlds in Horndeski gravity}
\author{F. A. Brito\inst{1,2} and Fabiano F. Santos\inst{1}
\thanks{\emph{fabiano.ffs23@gmail.com,fabrito2007@gmail.com}}}                     

%
%
\institute{$^{1}$Departamento de F\'\i sica,
Universidade Federal da Para\'\i ba, Caixa Postal 5008,
58051-970 Jo\~ ao Pessoa, Para\'\i ba,
Brazil\\
$^{2}$Departamento de F\'\i sica,
Universidade Federal de Campina Grande, Caixa Postal 10071,
58109-970  Campina Grande, Para\'\i ba,
Brazil}
\date{Received: date / Revised version: date}
%
\abstract{
In this paper we address the issue of finding braneworld solutions in a five-dimensional Horndeski gravity and the mechanism of gravity localization into the brane via ``almost massless modes'' for suitable values of the Horndeski parameters. We compute the corrections to the Newtonian potential and discuss the limit where four-dimensional gravity is recovered.
} 

\maketitle
\newpage

\newpage

\section{Introduction}

In recent years, theories of modified gravity have been of great interest \cite{Bahamonde:2019ipm,Bahamonde:2019shr,Clifton:2011jh}. In which the main motivation for studying such theories arose due to the recent observational cosmological data that suggest the need to introduce mysterious components, which may give us an idea of the past and the present universe, which serve as the basis for explaining a story consistent cosmic \cite{Rinaldi:2016oqp,Harko:2016xip,Bhattacharya:2016naa,Mukherjee:2017fqz}. When we probe the standard model, the extra dimensions suggest a way to probe physics beyond the standard model. Thus, the current detection of gravitational waves and electromagnetic signals, which result in the fusion of a binary system of compact objects, such as neutron stars, a consequence of such a phenomenon, is that the constraints of the geometry of extra dimensions arise for a space-time greater than $3+1$ dimensions \cite{Visinelli:2017bny}. Furthermore, when we carry out extensions of general relativity, we have that the candidates for the theory and phenomenology of unification are superstrings/M-theory \cite{Minamitsuji:2013vra}, which in this case these candidates are for a spacetime larger than four. Thus, we have that the idea behind extra-dimensional space leads to the development of the superstring/M-theory. In addition, if we consider an approach considering a dimension greater than four, considering an extra dimension is very small, we have that the context of string theory suggests to us a variety of ingredients in the extra-dimensional space, where we can build a cosmological and phenomenological model according to \cite{Minamitsuji:2013vra}. In fact, in this approach, cosmological models were built and are known as braneworld \cite{ArkaniHamed:1998rs,Antoniadis:1998ig,Randall:1999vf,Randall:1999ee} models, which state that we assume that our universe lives on a brane $3+1$-dimensional located on the border of deformed extradimensional space and that all fundamental interactions are confined to the brane.
 
However, investigations have recently been carried out with respect to Randall-Sundrum's well-known braneworld models \cite{Randall:1999vf,Randall:1999ee} using Horndeski's scalar-tensor theories. In this investigation carried out by \cite{Minamitsuji:2013vra} the author considered the model of John Lagrangian \cite{Bruneton:2012zk}, this is a specific model that is related to the F4 theories that were recently presented by \cite{Charmousis:2011bf,Charmousis:2011ea}, these theories are special subclasses of Horndeski. These coupling to the Einstein tensor is known as Horndeski gravity that was present in 1974, this theory is the most general scalar-tensor theory, in order to find such a theory of gravity Horndeski \cite{Zumalacarregui:2013pma,Cisterna:2017jmv,Rinaldi:2012vy,Heisenberg:2018vsk,horndeski74} firstly assumed the most general second-order Euler-Lagrange tensors derivable from a Lagrangian that is concomitant of a pseudo-Riemannian metric-tensor, a scalar field and their derivatives of arbitrary order in four-dimensional space --- although higher dimensional gravity was also discussed in \cite{horndeski74,Cisterna:2014nua}. The conclusion is that these Euler-Lagrange tensors may be obtained from a Lagrangian which is at most of the second order in the derivatives of the field functions.

In this paper we will focus on the study of braneworld solutions and localization of four-dimensional gravity in a Horndeski theory of gravity in five dimensions. This special truncation of the Horndeski gravity follows the same class of coupled scalar-tensor theories  such as the Brans-Dicke theory whose scalar field is non-minimally coupled to the Einstein-Hilbert term. But here the Einstein tensor couples non-minimally to the squared derivative of a dynamical scalar field. Several recent studies in black holes and gravitational waves have been considered in this context which imposes bounds on the parameters of the theory \cite{Feng:2015oea,Anabalon:2013oea,Cisterna:2014nua,Bettoni:2016mij}. Since braneworlds  can be understood as an effective low energy theory of superstrings which in turn can enclose the most general theories of gravity, it seems mandatory to investigate the phenomenon of  localization of four-dimensional gravity in the Horndeski theory.

Beyond of the Horndeski gravity that we will to consider to investigate the braneworld scenario other works in recent years have call attention, as for example \cite{Rosa:2022fhl,Silva:2022pfd,Afonso:2007zz,Moraes:2016gpe} where in this theories are considered braneworld models in generalized gravity theories with the action depends on a function of the Ricci scalar $R$ and the trace of the stress-energy tensor $T$. On the other hand, in our prescription for the John Lagrangian \cite{Bruneton:2012zk} we consider the coupling as $\eta G_{MN}\nabla^{M}\phi\nabla^{N}\phi$ where the $\eta$ parameter controls such coupling between the $G_{MN}$ Eintein tensor and the scalar field $\phi$ in five-dimensions. We explore the so-called scalar-tensor theories. For this, we introduce a first-order formalism to relate the scalar field of the Horndeski gravity sector with the  braneworld. 
 
The localization of gravity on a 3-brane \cite{Randall:1999vf} arises in the sense of being an alternative for compactification involving infinite extra dimensions. In the Randall-Sundrum scenarios \cite{Randall:1999vf,Randall:1999ee}, 3-branes are embedded into  $AdS_{5}$ $bulk$ space which is an ambient space developing 5-dimensional gravity with a negative cosmological constant ($\Lambda_{5}<0$) and infinitely thin sources of 3-branes composed of delta functions. In this setup there is a perfect fine-tuning between the brane tension and the cosmological constant $\Lambda_{5}$. Thus, the fine adjustment leads to a 4d brane in Minkowski $(M_{4})$ with a cosmological constant of four dimensions $\Lambda_{4d}=0$, in such a way that only the space $AdS_{5}$ is curved. In the limit of thick branes their profile are described by scalar fields, which is the main proposal here. The graviton zero-mode localized into the 3-brane is responsible for a localized 4d gravity, so that the correction for Newtonian potential due to Kaluza-Klein gravitons is highly suppressed in the case of low energies.  Even in the case of graviton `quasi zero mode' \cite{Karch:2000ct}, i.e, the case of massive gravity, the localization of gravity can still be achieved as metastable localization. We investigate the behavior of the coupling $\eta$ and their implications to the stability where in our work we will take an approach with respect to the \cite{DeWolfe:1999cp,Csaki:2000fc,Bazeia:2004yw}, in order to evaluate how the graviton will be localized in the brane with the use of Horndeski gravity. 

The paper is organized as follows. In Sec.~\ref{s1} we present the Horndeski gravity  and in Sec.~\ref{s2} we develop the first order formalism in five dimensions. In Sec.~\ref{s4} we compute numerical solutions. In Sec.~\ref{s5} we address the issue of the graviton fluctuations and in Sec.~\ref{s6} we compute the corrections to the Newtonian potential. Finally in Sec.~\ref{s7} we present our final comments.

\section{The Horndeski gravity with a scalar potential}
\label{s1}

In our present investigation we shall address the study of braneworlds in the framework of the Horndeski gravity \cite{horndeski74,Cisterna:2014nua,Feng:2015oea,Anabalon:2013oea} which action with a scalar potential reads
\begin{equation}
I[g_{MN},\phi]=\int{\sqrt{-g}d^{5}x\left[k(R-2\Lambda)-\frac{1}{2}(\alpha g_{MN}-\eta G_{MN})\nabla^{M}\phi\nabla^{N}\phi-V(\phi)\right]}\label{10}
\end{equation}
Note that we have a non-minimal scalar-tensor coupling where we can define a new field $\phi^{'}\equiv\Psi$. This field has dimension of $(mass)^{2}$ and the parameters $\alpha$ and $\eta$ control the strength of the kinetic couplings, $\alpha$ is dimensionless and $\eta$ has dimension of $(mass)^{-2}$. The Einstein field equations can still be given in the form 
\begin{equation}
G_{MN}+\Lambda g_{MN}=\frac{1}{2k}T_{MN}\label{11}
\end{equation}
where $T_{MN}=\alpha T^{(1)}_{MN}-g_{MN}V(\phi)+\eta T^{(2)}_{MN}$ with
\begin{equation}\begin{array}{rclrcl}
T^{(1)}_{MN} &=&\nabla_{M}\phi\nabla_{N}\phi-\frac{1}{2}g_{MN}\nabla_{P}\phi\nabla^{P}\phi\\
T^{(2)}_{MN}&=&\frac{1}{2}\nabla_{M}\phi\nabla_{N}\phi R-2\nabla_{P}\phi\nabla_{(M}\phi R^{P}_{N)}-\nabla^{P}\phi\nabla^{K}\phi R_{MPNK}\\
              &-&(\nabla_{M}\nabla^{P}\phi)(\nabla_{N}\nabla_{P}\phi)+(\nabla_{M}\nabla_{N}\phi)\square\phi+\frac{1}{2}G_{MN}(\nabla\phi)^{2}\\
							&-& g_{MN}\left[-\frac{1}{2}(\nabla^{P}\nabla^{K}\phi)(\nabla_{P}\nabla_{K}\phi)+\frac{1}{2}(\square\phi)^{2}-(\nabla_{P}\phi\nabla_{K}\phi)R^{PK}\right]\label{g}
\end{array}\end{equation}
and the scalar field equation is 
\begin{equation}
\nabla_{M}[(\alpha g^{MN}-\eta G^{MN})\nabla_{N}\phi]=V_{\phi}\label{12}
\end{equation}
The metric {\it Ansatz} to be studied in five dimensions is of the form
\begin{eqnarray}
ds^{2}&=&g_{MN}dx^Mdx^N\nonumber\\
&=&e^{2A(r)}g_{\mu\nu}dx^{\mu}dx^{\nu}-dr^{2}\label{13}
\end{eqnarray}
where the latin indices $M,N=$ $0$, $1$, $2$, $3$ and $5$ run on the bulk and the greek indices $\mu,\nu=$ $0$, $1$, $2$ and $3$ run along the braneworld. 

\section{Equations of motion}
\label{s2}

In this section we analyze the equations of motion considering the metric (\ref{13}) we have the following Einstein equation components involving the scalar potential. First, from equation (\ref{11}) the $rr$-component is given by
\begin{equation}
2V(\phi)+4k(\Lambda+6A^{'2})-\Psi^{2}(r)[\alpha-18\eta A^{'2}]=0\label{14},
\end{equation}
where $\Psi(r)=\phi^{'}(r)$. This equation can be rewritten as follows
\begin{equation}
\Psi(r)=\pm\sqrt{\frac{2V(\phi)+4k\Lambda+24kA^{'2}(r)}{\alpha-18\eta A^{'2}(r)}}\label{141}
\end{equation}
Now taking the relationship between $A(r)$ and the superpotential $W(\phi)$ through the first-order differential equation 
\begin{equation}\label{ApW}
A^{'}(r)=-\frac13W(\phi)
\end{equation}
and redefining $24k\equiv b$ we can rewrite the equation (\ref{141}) in the form
\begin{equation}
\Psi(r)=\pm\sqrt{\frac{2V(\phi)+\frac{b\Lambda}{6}+\frac{b}{9}W^{2}(\phi)}{(\alpha-2\eta W^{2}(\phi))}}\label{142}
\end{equation}
For a potential of the form
\begin{equation}
V(\phi)=-\frac{b}{18}W^{2}(\phi)+\frac{c^{2}W^{2}_{\phi}(\phi)}{2}(\alpha-2\eta W^{2}(\phi))\label{2v}
\end{equation}
and taking $\Lambda=0$, for simplicity, we find
\begin{equation}
\Psi(r)=\pm{cW_{\phi}(\phi)}\label{3v}
\end{equation}
Another equation from the Einstein equations can be found by combining the $tt$-component with $xx$, $yy$ or $zz$-components to find
\begin{equation}
2V(\phi)+6\eta\Psi(r)\Psi^{'}(r)A^{'}(r)+4k[\Lambda+6A^{'2}(r)+3A^{''}(r)]+\Psi^{2}(r)[\alpha+6\eta A^{'2}(r)+3\eta A^{''}(r)]=0\label{15}
\end{equation}
A third equation describing the scalar field dynamics comes from the equation (\ref{12})
\begin{equation}
V_{\phi}(\phi)+4A^{'}\phi^{'}(r)[-\alpha+6\eta A^{'2}(r)+3\eta A^{''}(r)]-(\alpha-6\eta A^{'2}(r))\phi^{''}=0\label{16}
\end{equation}
Using the following relationships $\phi^{'}=cW_{\phi}$, $A^{''}=-cW^{2}_{\phi}/3$ we can write these equations as
\begin{equation}
2cWW_{\phi\phi}+cW^{2}_{\phi}+\frac{4}{3}W^{2}-\frac{(2\alpha-b/6c)}{\eta}=0\label{15.1}
\end{equation}
This implies that the equations (\ref{14}), (\ref{15}) and (\ref{16}) are consistently satisfied by equation  (\ref{ApW}) and (\ref{3v}) under the nontrivial condition (\ref{15.1}) on the superpotential $W(\phi)$. The equation (\ref{15.1}) can solve numerically as we show in next section.

\section{Numerical solutions}
\label{s4}

\begin{figure}[!ht]
\begin{center}
\includegraphics[scale=0.3]{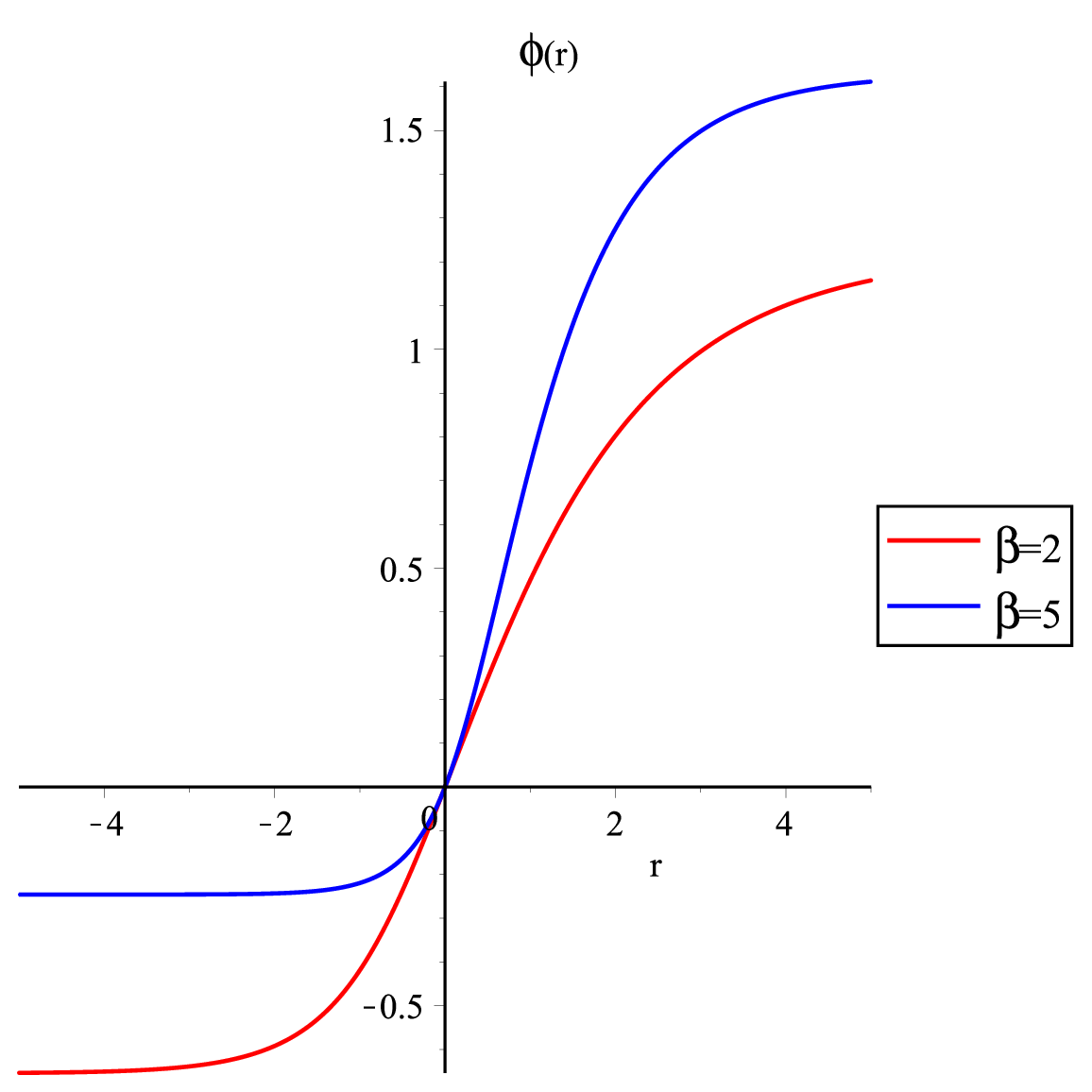}
\includegraphics[scale=0.3]{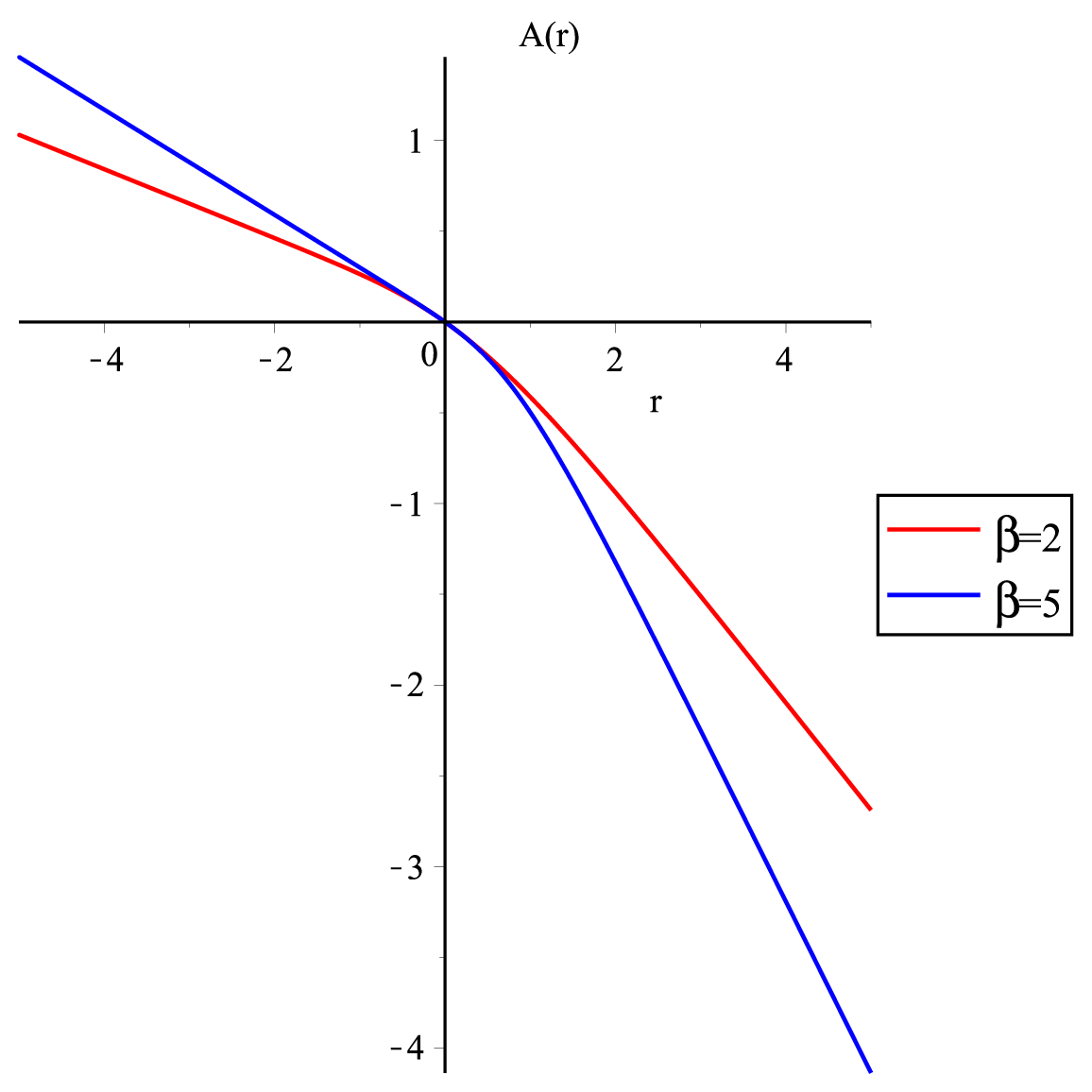}
\includegraphics[scale=0.3]{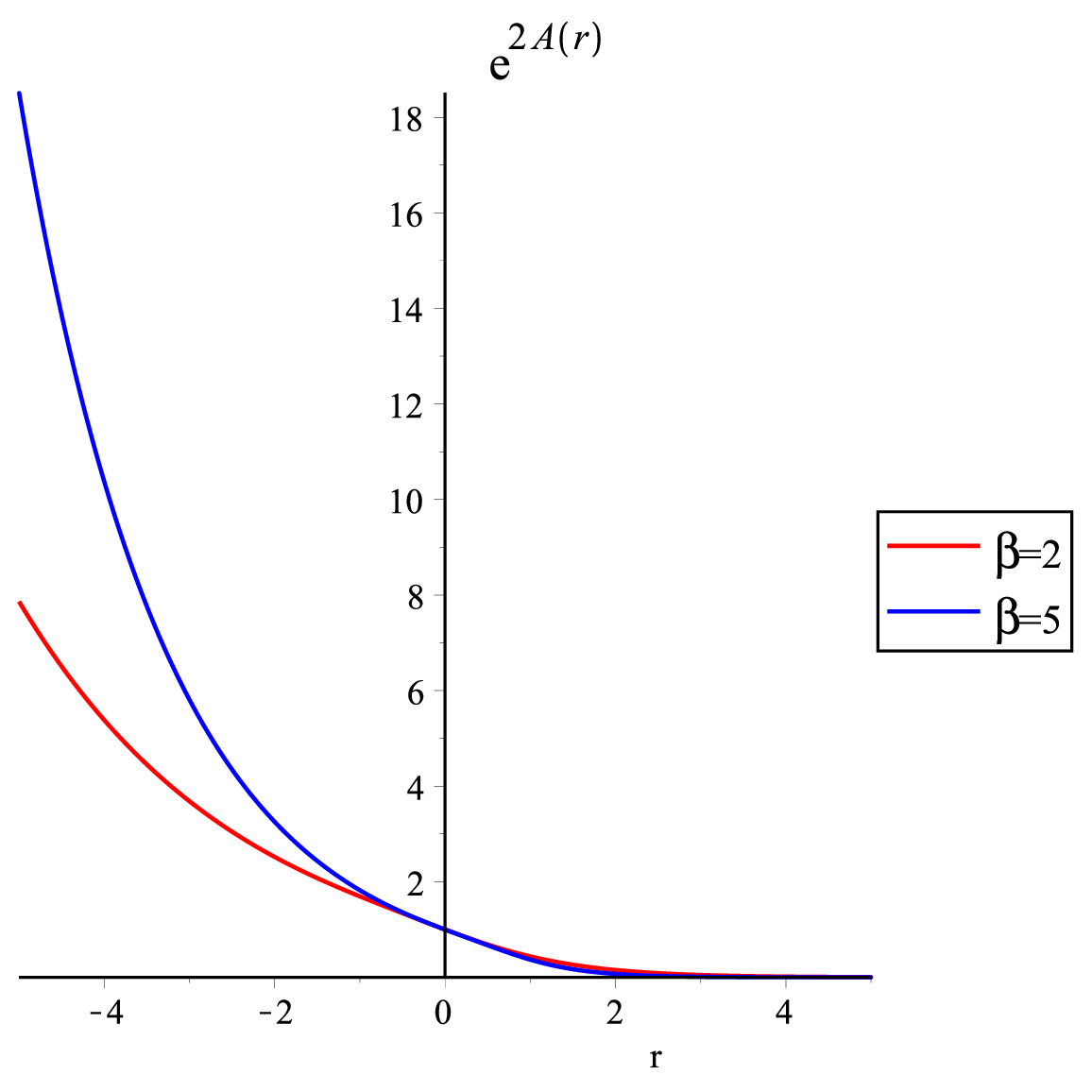}
\caption{The figure shows the kinks $\phi(r)$ (top-left) and $A(r)$ (top-right) for $\beta=2$ (red) and $\beta=5$ (blue) for $c=1/2$. In the bottom we show the warp factor $\exp{(2A(r))}$ signalizing the existence of an asymmetric brane.}
\label{horn-nums-figs}
\end{center}
\end{figure}

In the examples above we were able to find explicit solutions in a restrict regime of value of $\alpha$ and $\eta$. However, the pair of first-order equations (\ref{ApW}) and (\ref{3v}) can be solved numerically for a broader range of values as long as we assume appropriate boundary conditions. In Fig.~\ref{horn-nums-figs} we show the behavior of the kink profile and geometry associated with the braneworld solutions for $\beta=2$ and $5$ for $c=1/2$, which means $2(\alpha-1)/\eta$=2 and 5, respectively, that for $\alpha=3/2$ we have $\eta=0.5$ (red curve) and $\eta=0.2$ (blue curve). This is the regime of small $\eta$, which was not addressed previously in the non-homogeneous limit of Eq.~(\ref{15.1}). The warp factor signalizes the existence of an asymmetric brane, which in general is not able to localize graviton zero mode, but at least one may find metastable gravity. Lower values of $\eta<1$ seems to force brane asymmetry. The boundary conditions used was the following: $W(0)=1$, $W'(0)=1$, $\phi(0)=0$ and $A(0)=0$.

\section{Equation for fluctuations}
\label{s5}

\begin{figure}[!ht]
\begin{center}
\includegraphics[scale=0.3]{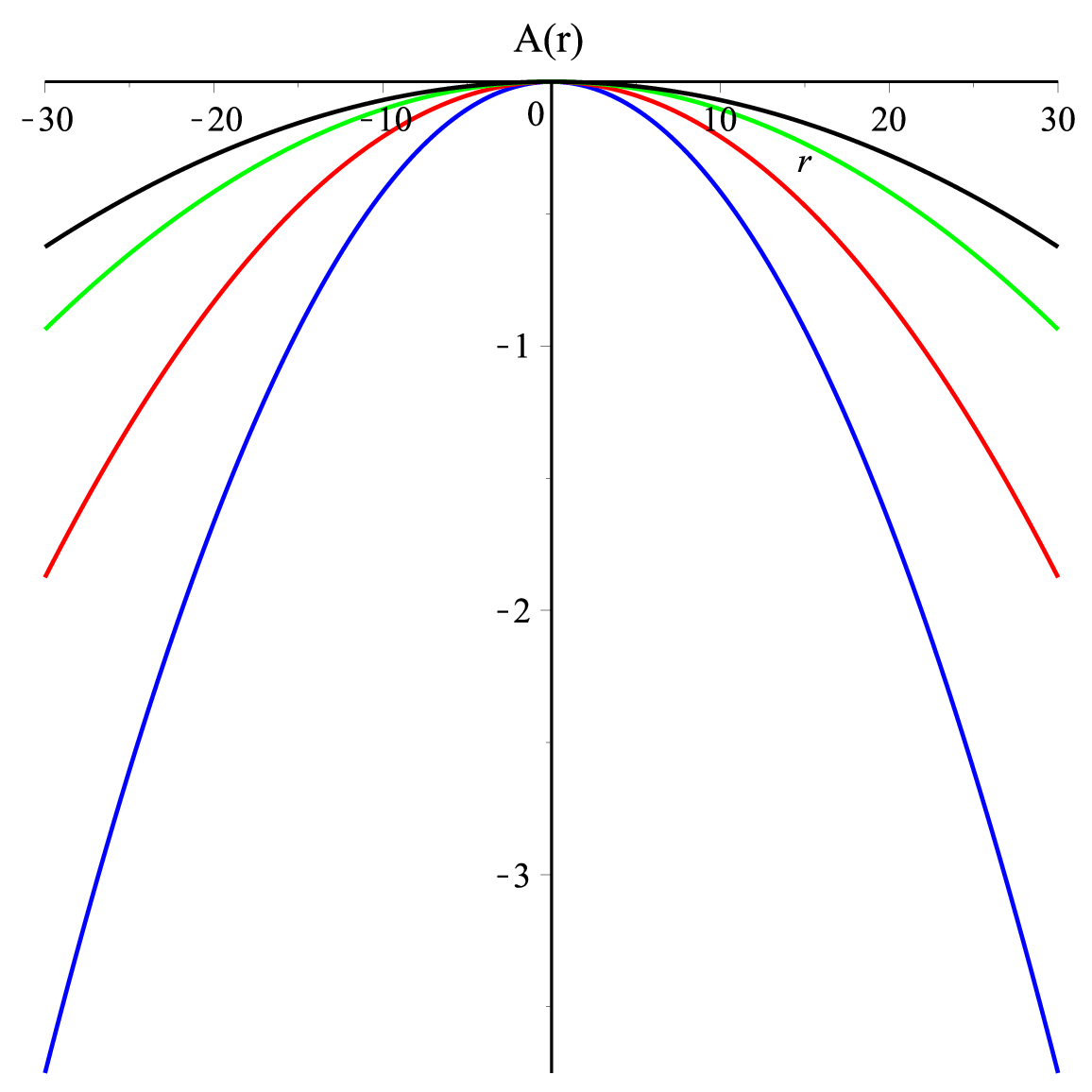}
\caption{The figure shows $A(r)$ for $q=\sigma^{2}/6$ and $\alpha=1.5$, $\sigma=\sqrt{4(\alpha-1)/\eta}$, $\eta=10$-blue, $\eta=20$-red, $\eta=40$-green and $\eta=60$-black.}
\label{A-r}
\end{center}
\end{figure}

Let us now compute  the linearization of the Einstein equations by considering the following perturbations $\bar{g}_{MN}=g_{MN}+h_{MN}$ in the axial gauge $h_{M 5}=0$, where $M$ is a bulk index. Thus, the fluctuation in the metric (\ref{13}) can be  written in the form
\begin{eqnarray}
ds^{2}=e^{2A(z)}((g_{\mu\nu}+h_{\mu\nu})dx^{\mu}dx^{\nu}-dz^{2})
\end{eqnarray}
Now, performing the following first-order perturbations $\delta^{(1)}g_{\mu\nu}=h_{\mu\nu}$, where $h_{\mu\nu}$ is the transverse and traceless (TT) tensor perturbation, that is, $\eta^{\mu\alpha}=\partial_{\alpha}h_{\mu\nu}=0$ and $h\equiv\eta^{\mu\nu}h_{\mu\nu}=0$ \cite{DeWolfe:1999cp,Fu:2019xtx}. However, considering the (TT), we can write 

\begin{eqnarray}
&&C(z)\partial^{2}_{z}h_{\mu\nu}+D(z)\partial_{z}h_{\mu\nu}+\Box_{4d}h_{\mu\nu}=0\label{T1}\\
&&C(z)=\frac{1-\eta e^{-2A}\phi^{'2}}{1+\eta e^{-2A}\phi^{'2}}\label{T2}\\
&&D(z)=\frac{3A^{'}-\eta e^{-2A}A^{'}\phi^{'2}-2\eta e^{-2A}\phi^{'2}\phi^{''}}{1+\eta e^{-2A}\phi^{'2}}\label{T3}
\end{eqnarray}
Now, considering the following coordinate transformation $dz=\sqrt{C}d\omega$, we can write for the equation (\ref{T1}):
\begin{eqnarray}
\partial^{2}_{\omega}h_{\mu\nu}+\left(\frac{D}{\sqrt{C}}-\frac{\partial_{\omega}C}{2C}\right)\partial_{\omega}h_{\mu\nu}+\Box_{4d}h_{\mu\nu}=0.\label{T4}
\end{eqnarray} 
where by the decomposition $h_{\mu\nu}(x,\omega)=\epsilon_{\mu\nu}(x)e^{-ipx}H(\omega)$ with $p^{2}=-m^{2}$, we have
\begin{eqnarray}
&&\partial^{2}_{\omega}H(\omega)+Q(\omega)\partial_{\omega}H(\omega)+m^{2}H(\omega)=0,\label{T5}\\
&&Q(\omega)=\frac{D}{\sqrt{C}}-\frac{\partial_{\omega}C}{2C}\label{T6},
\end{eqnarray}
However, we can simplify the equation (\ref{T5}), by redefining $H(\omega)=G(\omega)\psi(\omega)$ with $G(\omega)=\exp\left(-\frac{1}{2}\int{Q(\omega)d\omega}\right)$, we can compute the Schrödinger-like equation:
\begin{eqnarray}
&&-\partial^{2}_{\omega}\psi(\omega)+U(\omega)\psi(\omega)=m^{2}\psi(\omega),\label{T7}\\
&&U(\omega)=\frac{Q^{2}}{4}+\frac{\partial_{\omega}Q}{2}\label{T8},
\end{eqnarray}
This is an unusual potential as compared with those in the literature \cite{DeWolfe:1999cp,Csaki:2000fc,Bazeia:2004yw,Karch:2000ct}. However, one can easily recover the usual case as $\alpha=1$ and $\eta=0$, otherwise we have the potential derived from Horndeski gravity. The equation (\ref{T7}) can be factorize as
\begin{eqnarray}
\left(\partial_{\omega}+\frac{Q}{2}\right)\left(-\partial_{\omega}+\frac{Q}{2}\right)\psi(\omega)=m^{2}\psi(\omega),\label{T9}
\end{eqnarray}
where we can found that there is no tachyon state, that is, $m^{2}\geq 0$. Furthermore, we can found the zero mode for solve the equation (\ref{T7}) by setting $m=0$, we have
\begin{eqnarray}
\psi_{0}=C_{0}\exp\left(\frac{1}{2}\int{Qd\omega}\right)=C_{0}\exp\left(\frac{1}{2}\int{\frac{Qdz}{\sqrt{C}}}\right),\label{T10}
\end{eqnarray}
where $C_{0}$ is a normalization constant. Thus, we have that the normalization condition associated to the graviton zero mode is given by
\begin{eqnarray}
\int{\psi_{0}(\omega)d\omega}<\infty.\label{T11}
\end{eqnarray}
However, to analyze the ``almost massless modes'', we need of a analytical solutions. Thus, recent investigation shown that mechanism of gravity localization described by the gaussian warp factor \cite{Quiros:2012bh,Llatas:2001jj} lead to the local harmonic approximation. Based in this assumption in our work we analyze the graviton fluctuations, so looking for the equation (\ref{15.1}) we present a way of found a analytical gaussian warp factor as presented in the following cases:

i) For $\alpha=b/12c$ with $c=1/2\eta$ that imply in $\alpha=\eta/(4\pi G)$, this relation between the parameters $\alpha$ and $\eta$ is very familiar when analytical example are considered in Horndeski gravity for the black branes solutions \cite{Bravo-Gaete:2013dca}. However, we can see that $\alpha=\eta/(4\pi G)$ represent a non-degenerate point. Note this choice is dimensionally correct since $[\eta]=1/mass^{2}$ and $[b]=[6/4\pi G]=mass^{2}$, Eq.~(\ref{15.1}) becomes a homogeneous differential equation that gives the solution
\begin{equation}
W(\phi)=\left(\frac{1}{2}\right)^{2/3}\left(-C_{1}\sin\left(\frac{\phi}{\sqrt{c}}\right)-C_{2}\cos\left(\frac{\phi}{\sqrt{c}}\right)\right)^{2/3}
\end{equation}
With this superpotential we can easily solve the first-order differential equations by implementing a numerical Runge-Kutta method. We come back to this approach shortly.
However, by expanding this superpotential for $\eta\ll1$ we have
\begin{equation}
W(\phi)=\gamma+\sigma\phi
\end{equation}
where the $\gamma=(-C_{2}/2)^{2/3}$ and $\sigma=2C_{1}(-C_{2}/2)^{2/3}/3\sqrt{c}C_{2}$ are constants. Note that for $C_{2}$ very small we can suppress $\gamma$. Using $\phi^{'}=cW_{\phi}$ we can write 
\begin{equation}
W(r)=c\sigma^{2}(r-r_{0})
\end{equation}
where $\phi=c\sigma r$. Thus, we find 
\begin{equation}
A(r)=-\frac{\sigma^{2}}{12\eta}(r-r_{0})^{2}
\end{equation}
or for $\sigma^{2}/6\eta=q$ and $r_{0}=0$ we can simply  write
\begin{equation}
A(r)=-\frac{q}{2}r^{2}\label{WARP}
\end{equation} 

ii) For $\beta=2(\alpha-1)/\eta$ with $b=6, c=1/2$,
Eq.~(\ref{15.1}) is not a homogeneous differential equation but can still be approximately solved by the linear solution
\begin{equation}
W(\phi)=\sigma\phi,\qquad \sigma=\sqrt{2\beta}
\end{equation}
where $\sigma\ll1$ for very large $\eta$. Thus by using Eqs.~(\ref{ApW}) and (\ref{3v}) we have
\begin{equation}
\phi(r)=\frac12\sigma r, \qquad A(r)=-\frac{q}{2}r^{2}
\end{equation} 
where $q=\sigma^2/6$. $A(r)$ is depicted in Fig.~\ref{A-r} for distinct values of $\eta$ --- similar type of geometry has been already considered in confining AdS/QCD \cite{Andreev:2006ct}. The scalar solutions discussed in the two examples above can be understood as very thick `kinks' which have very small slope. We shall show numerically an evidence that the slope of the kink type solution of the exact differential equations indeed tends to decrease as $\eta$ increases.

\begin{figure}[!ht]
\begin{center}
\includegraphics[scale=0.4]{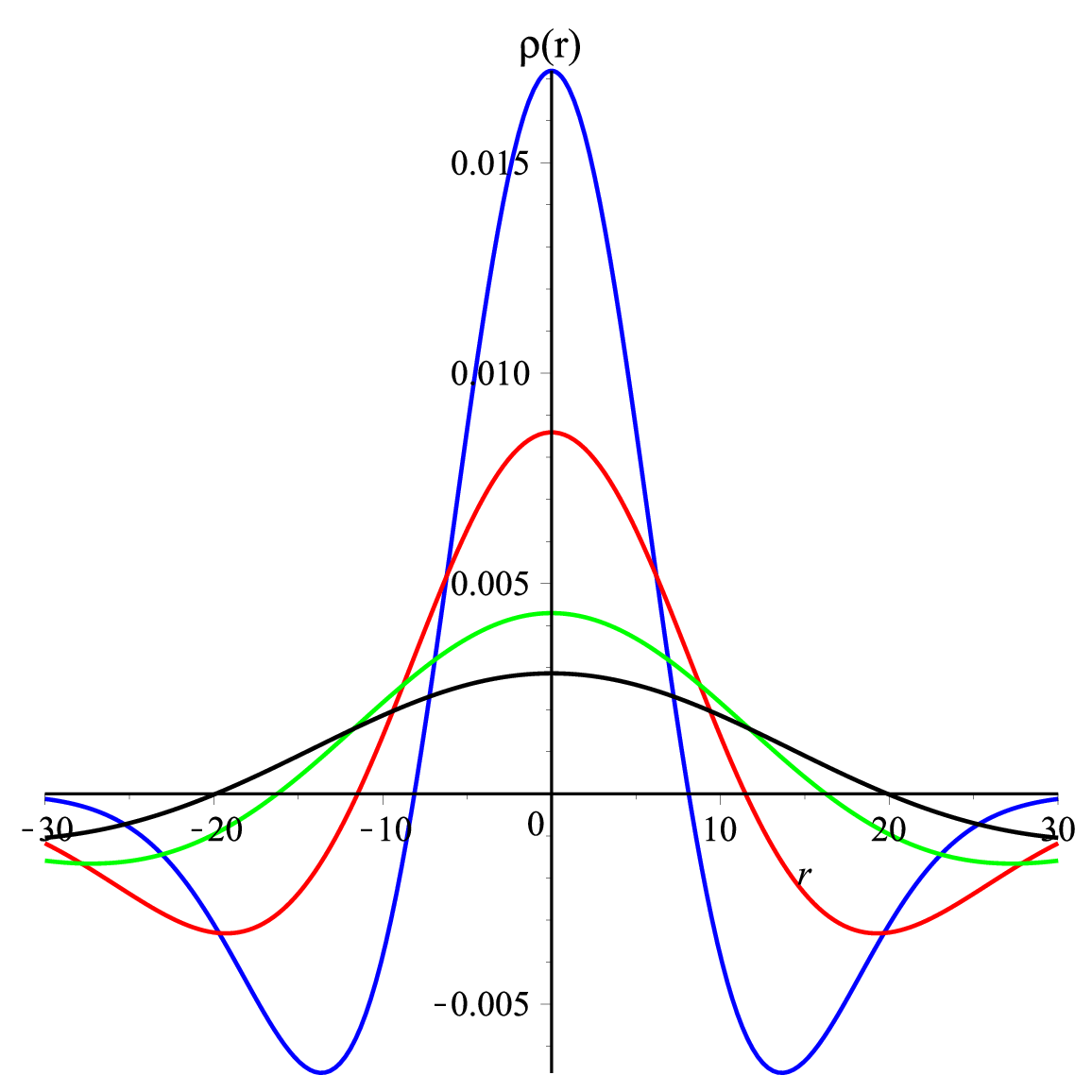}
\caption{The energy density $\rho(r)$ for small values of $q=\sigma^{2}/6$, $\alpha=1.5$, $\sigma=\sqrt{4(\alpha-1)/\eta}$, $\eta=10$-blue, $\eta=20$-red, $\eta=40$-green and $\eta=60$-black.}\label{A6.2}
\end{center}
\end{figure}

The energy density is given by
\begin{eqnarray}
\rho(r)=-e^{2A}\left(\frac{3\alpha A^{''}}{2}+3A^{'2}\right)-\frac{9\eta e^{2A}}{4}(A^{'}A^{'''}-4A^{'2}A^{''}+A^{''2})
\end{eqnarray}
and for such solutions the energy density is depicted in Fig.~\ref{A6.2} for several large values of $\eta$.
Note that in Fig.~\ref{A6.2} when the parameter $\eta$  increases, we can see that the warp factor Eqs.~(\ref{WARP}) also becomes more and more unconcentrated. On the other hand, for the $\eta=10$, we have a more and more concentrated energy density.
To solve the equation (\ref{T7}) let us first recall the above transformation of variables between $r$ and $z$ by using $z(r)=\int{e^{-A(r)}}dr$ such that we have
\begin{equation}
z(r)=\sqrt{\frac{\pi}{2q}}{\rm erf}\left(\sqrt{\frac{q}{2}}r\right)=\frac{2}{\sqrt{2q}}\sum^{\infty}_{k=0}\frac{(-1)^{k}\left(\sqrt{\frac{q}{2}}r\right)^{2k+1}}{k!(2k+1)}\label{T12}
\end{equation}
whose first terms are  
\begin{equation}
z(r)=\frac{2}{\sqrt{2q}}\left(\sqrt{\frac{q}{2}}r-\frac{(\sqrt{\frac{q}{2}}r)^{3}}{3}+\frac{(\sqrt{\frac{q}{2}}r)r^{5}}{10}+...\right)\label{T13}
\end{equation}
Now recalling that for $\eta\gg1$  implies $q\ll1$, for the second example above, then taking the leading term for this approximation we find $z=r$, which allows us to write
\begin{equation}
A(z)=-\frac{q}{2}z^{2}\label{T14}
\end{equation}
Using this geometry we can be write

\begin{eqnarray}
&&C(z)=\frac{1-(\alpha-1)e^{qz^{2}}}{1+(\alpha-1)e^{qz^{2}}}\label{T15}\\
&&D(z)=-3qzC(z)\label{T16}
\end{eqnarray}
As $q\ll1$ we can expand (\ref{T15}) up to first order and obtain $C(z)=\frac{2-\alpha}{\alpha}+\vartheta(z)$, which is a constant. Now, through the transformation $dz=\sqrt{C}d\omega$, we have $\omega=\sqrt{C(z)}z$. Thus, we can write that
\begin{eqnarray}
U(\omega)=\frac{9q^{2}\omega^{2}}{4}-\frac{q\sqrt{C}}{2}\label{T71},
\end{eqnarray}
As $\eta$ is sufficiently large, $q$ is tiny enough. Thus, we can define that
\begin{eqnarray}
V(\omega)=M^{2}\omega^{2}\label{00p}, \qquad 
M\equiv\  \frac{3q}{2}\label{VW}
\end{eqnarray}
such a way we can now rewrite the equation (\ref{T71}) in the form
\begin{equation}
-\partial^{2}_{\omega}\psi(\omega)+M^{2}\omega^{2}\psi(\omega)=E\psi(\omega)\label{1f}
\end{equation}
Note that this equation is a Schroedinger-like problem for a quantum harmonic oscillator whose solution is well-known in the quantum mechanics literature. The equation (\ref{1f}) can be written as an eigenvalue equation, i.e., $H\psi=E\psi$ with the Hamiltonian
\begin{equation}
H=-\partial^{2}_{\omega}+M^{2}\omega^{2}
\end{equation}
We can define the following  ``annihilation" and ``creation" operators written as follows
\begin{equation}
Q=\frac{1}{\sqrt{M}}(-\partial_{\omega}-M\omega)\quad \mbox{and}\quad Q^{+}=\frac{1}{\sqrt{M}}(\partial_{\omega}-M\omega)
\end{equation}
which implies in a Hamiltonian written in terms of  such operators in the form
\begin{equation}
H=M(N+1)
\end{equation}
where $N=Q^{+}Q$. Note that the energies for this ``oscillator" are of the form $E=M(n+1)$ which implies $m^{2}_{n}=M(n+1)$. Thus, the graviton spectrum reads
\begin{equation}
m^{2}_{n}=M(n+1)+(1-\alpha)\sqrt{\frac{2-\alpha}{\alpha}}\label{1g}
\end{equation}
For AdS/QCD theories that follows a similar linear glueball spectrum see, e.g. \cite{Brower:2000rp,Karch:2006pv}.
For $n=0$ we have that $m^{2}_{0}=M+(1-\alpha)\sqrt{\frac{2-\alpha}{\alpha}}$. In the limit of large $\eta$, we have two values for $\alpha$, which are $\alpha=1,2$ and one achieves $m^{2}_{0}\rightarrow 0$ that plays the role of a ``quasi-zero mode''  which is responsible for the $4d$ gravity on the brane. Let us now evaluate the wave-function of the quasi-zero mode through the equation $Q\psi_{0}(\omega)=0$ which implies
\begin{equation}
\psi_{0}(\omega)=Ce^{\frac{-M\omega^{2}}{2}}\label{1j}
\end{equation}
\begin{figure}[!ht]
\begin{center}
\includegraphics[scale=0.3]{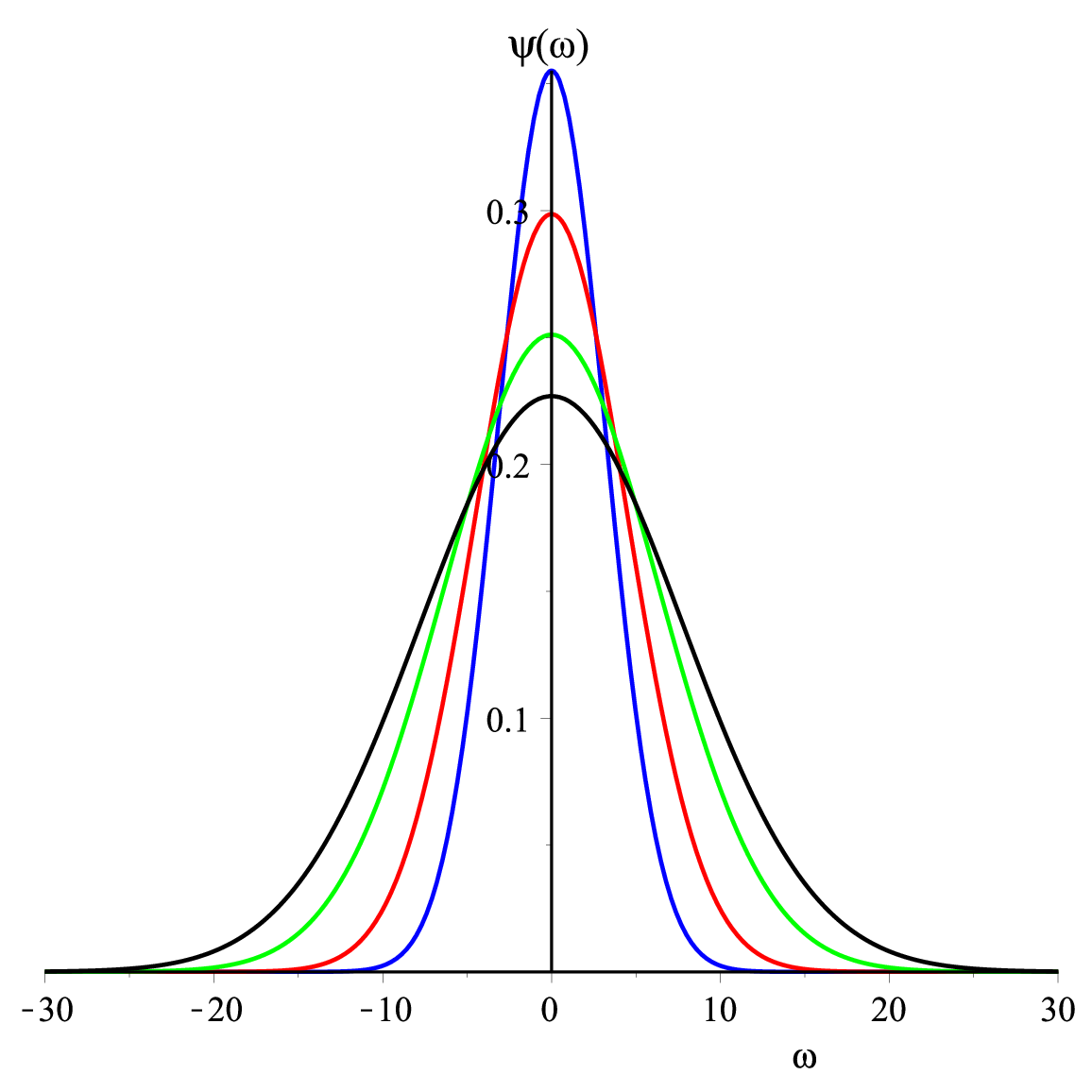}
\caption{The wave-function $\psi_{0}(\omega)$ for values of $q=\sigma^{2}/6$, $\alpha=2$, $\sigma=\sqrt{4(\alpha-1)/\eta}$, $\eta=10$-blue, $\eta=20$-red, $\eta=40$-green and $\eta=60$-black.}
\end{center}
\end{figure}
The eigenfunctions of the tower of higher massive modes are given by the solutions of equation (\ref{1f}) which read
\begin{equation}
\psi_{n}(\omega)=\frac{C}{\sqrt{2^{n}n!}}\bar{H}_{n}(M\omega)e^{\frac{-M\omega^{2}}{2}}
\end{equation}
\begin{figure}[!ht]
\begin{center}
\includegraphics[scale=0.3]{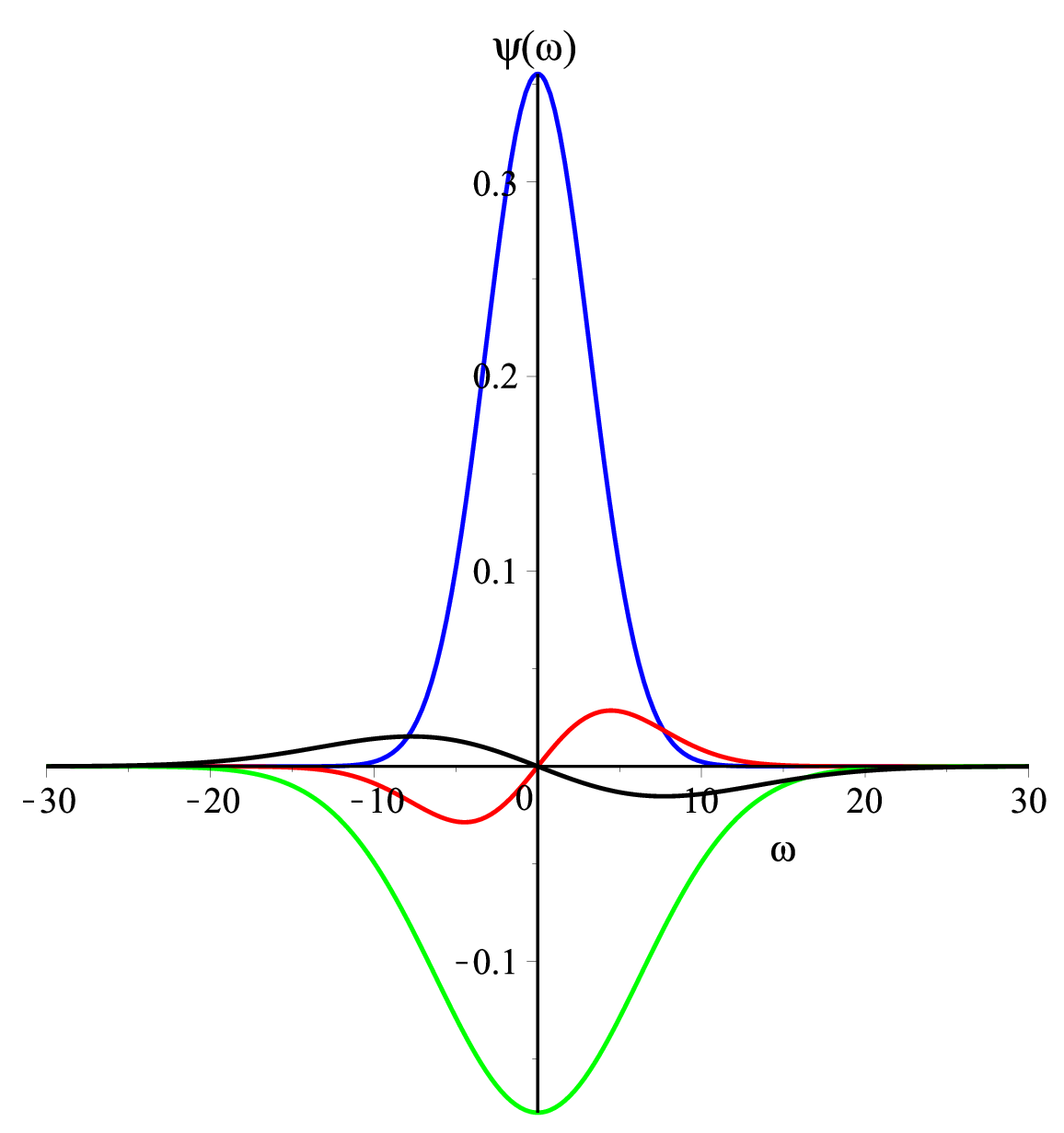}
\caption{The wavefunctions $\psi_{n}(\omega)$ for values of $q=\sigma^{2}/6$ due to the geometry given with the values of $\sigma=\sqrt{4(\alpha-1)/\eta}$, $\alpha=2$, $n=0$/$\eta=10$-blue, $n=1$/$\eta=20$-red, $n=2$/$\eta=40$-green and $n=3$/$\eta=60$-black.}
\label{planohwkhz}
\end{center}
\end{figure}
where the functions $\bar{H}_{n}(M\omega)$ are the well-known Hermite polynomials and $C=\left(\frac{M}{\pi}\right)^{1/4}$.With the computations of the graviton spectrum, we can analyze the Newtonian potential in the next section. However, under an appropriate choice of the Horndeski parameters, this change of shape in the potential implies that the graviton zero mode $\psi_{0}$ is localized in a wider region around the origin. Given the finiteness of the integral of $\psi_{0}$ over $z$, one concludes that braneworld is stable for fluctuations in the metric.

\section{Correction to Newtonian potential}
\label{s6}

The effective Newtonian potential in braneworlds scenarios can be found as in the form \cite{Csaki:2000fc}
\begin{eqnarray}
V(r)=\psi^{2}_{0}(0)\frac{e^{-m_{0}r}}{r}+\sum_{n}\psi^{2}_{n}(0)\frac{e^{-m_{n}r}}{r}
\end{eqnarray}
Notice that the Hermite polynomials at $\omega=0$ contribute to the potential only for even modes, that means
\begin{eqnarray}
\psi^{2}_{0}(0)&=&C^{2},\\
\psi^{2}_{n}(0)&=&\frac{C^{2}\bar{H}^{2}_{n}(0)}{2^{n}n!}=\frac{\pi2^{n}C^{2}}{\Gamma\left(\frac{1-n}{2}\right)^{2}\Gamma(n+1)}
\end{eqnarray}
As $\eta$ is very large $q\rightarrow 0$ and considering the two values of $\alpha=1,2$ the graviton mass spectrum approaches a continuum and we can substitute the sum for an integral. Thus, we can write  \cite{Schwartz:2000ip}

\begin{eqnarray}
V(r)&\approx&C^{2}\frac{e^{-\sqrt{M}r}}{r}+\frac{C^{2}}{r}\int{\frac{\pi2^{n}}{\Gamma\left(\frac{1-n}{2}\right)^{2}\Gamma(n+1)}e^{-\sqrt{M(n+1)}r}dn}\\
    &\approx&\frac{1}{r}\sqrt{\frac{M}{\pi}}\left(e^{-\sqrt{M}r}+\pi\int{\frac{2^{n}}{\Gamma\left(\frac{1-n}{2}\right)^{2}\Gamma(n+1)}e^{-\sqrt{M(n+1)}r}dn}\right)
\end{eqnarray}
Let us now make a change of variables as in the form $x=M(n+1)$ to which we have
\begin{eqnarray}
V(r)\approx\frac{1}{r}\sqrt{\frac{M}{\pi}}\left(e^{-\sqrt{M}r}+\frac{\pi}{2M}\int{\frac{2^{\frac{x}{M}}}{\Gamma\left(1-\frac{x}{2M}\right)^{2}\Gamma\left(\frac{x}{M}\right)}e^{-x^{1/2}r}dx}\right)
\end{eqnarray}
In the limit of highly excited modes, that is $x/M\rightarrow \infty$, the arguments of the $\Gamma$ functions become large and we can use the following approximation 
\begin{eqnarray}
\frac{2^{\frac{x}{M}}}{\Gamma\left(1-\frac{x}{2M}\right)^{2}\Gamma\left(\frac{x}{M}\right)}\approx{\frac{2\sqrt{2}}{\pi^{3/2}}}\frac{\sqrt{M}}{\sqrt{x}\left(1-\cos^{2}\left(\frac{\pi}{2}(n+1)\right)\right)^{-1}}
\end{eqnarray}
As we have previously mentioned (and also easily checked from Fig.~\ref{planohwkhz}) the Hermite polynomials at $z=0$ do not contribute to the potential for odd modes since $\bar{H}_{2n+1}(0)=0$, thus the cosine vanishes because its argument is odd multiples of $\pi/2$ for any even number $n$. As such, we can write   
\begin{eqnarray}
V(r)\approx\frac{1}{r}\sqrt{\frac{M}{\pi}}\left(e^{-\sqrt{M}r}+\frac{\sqrt{2}}{\sqrt{\pi M}}\int^{\infty}_{1/r_c^2}{\frac{1}{\sqrt{x}}e^{-x^{1/2}r}dx}\right)
\end{eqnarray}
where $r_c$ is a crossover scale at large energies. We can now perform the integral in the second term to find 
\begin{eqnarray}
V(r)\approx\frac{1}{r}\sqrt{\frac{M}{\pi}}\left(e^{-\sqrt{M}r}+\frac{2\sqrt{2}}{\sqrt{\pi M}}\frac{e^{-\frac{r}{r_c}}}{r}\right)
\end{eqnarray}
The result of the integral when calculated is real for the condition $\Re[r\sqrt{M}> 0]$ or $\Re(r)>0$. Recalling that $M$ is very small, the exponential can still be approximated to unity. Thus, as $r\ll r_c$ we obtain the final form of the corrected Newtonian potential as follows
\begin{eqnarray}
V(r)\approx\sqrt{\frac{M}{\pi}}\left(\frac{1}{r}+\frac{2\sqrt{2}}{r^{2}\sqrt{\pi M}}\right)\label{g15}
\end{eqnarray}
However, for $r\gg r_c$ one can recover the four-dimensional gravity. In addition, the computation of the weakly excited modes has shown to be able to correct the potential with an extra term that goes with $1/r^5$, which is easily suppressed at large distances in comparison with the $1/r$ term, but strongly dominant at small distances.  

\section{Conclusions}
\label{s7}

In this paper we have addressed the problem of localizing gravity in braneworlds solutions found in Horndeski gravity. Such solutions were found through the formalism applied to Horndeski gravity as well as to $f(R)$, $f(R,T)$ theories \cite{Afonso:2007zz,Moraes:2016gpe}. The idea behind from this formalism concerns in reducing the equations of motion to first-order equations, which simplifies the solution of the problem from both analytical and numerical perspective. In our prescription we show that numerical solutions was very important to check that the warp factor signalizes the existence of an asymmetric brane and with the Horndeski gravity fluctuations we provide the graviton zero mode, that is, we find metastable gravity. The $\eta$ parameter that deviates such theory  of gravity in relation to Einstein gravity controls the localization of four-dimensional gravity in a non-trivial way as can be easily checked directly in the induced Newtonian potential. 

Interestingly for sufficiently large $\eta$ the four-dimensional gravity is safely localized on the brane. This is also the regime where one can easily find explicit braneworlds solutions. It was in this regime that we restricted ourselves to the analysis of the Newtonian potential. Further studies for arbitrary values of $\eta$ should be addressed elsewhere. However, for $r\gg r_c$ one can recover the four-dimensional gravity. In addition, the computation of the weakly excited modes has shown to be able to correct the potential with an extra term that goes with $1/r^5$, which is easily suppressed at large distances in comparison with the $1/r$ term, but strongly dominant at small distances. We can see that the  explicit braneworlds solution in Horndeski gravity is a normalizable bound zeromass gravitational state whose wave-function shapes the form of the brane in the extra-space. The continuum of massive modes produces only very small (negligible) corrections to the Newtonian gravitational potential which fall-off very quickly as $1/r^5$, but the term $1/r$ is strongly dominant at small distances. This is expected since the analog quantum mechanical potential in Eq.$\sim$(\ref{VW}) $V( \omega\to\pm\infty)>0$, it uncontrollably grows up at large $|\omega|$. In our case the excited states are separated by a gap from the ground state that are controlled by the Horndeski paramters. Thus, in our set-up this may be understood in the following way. The massive modes are localized away from the brane, while the lighter ones are farther away while the heavier states are closer to the brane. 

However, in the future we think that it will be of special interest to explore the cosmological solutions, domain wall solutions and complexity through the first order formalism in the Horndeski gravity.


We would like to thank CNPq and CAPES for partial financial support. We also thank  Cristi\'an Erices for useful discussions in the early stages of this work.

\end{document}